\documentclass[11pt, a4paper]{article}
\usepackage{amssymb, geometry, jheppub}   

\title{Supersymmetric, fermionic solutions in three-dimensional supergravity}

\author{N. Houston}
\emailAdd{nhouston@itp.ac.cn}
\affiliation{Kavli Institute for Theoretical Physics China (KITPC) \& Key Laboratory of Theoretical Physics, Institute of Theoretical Physics, Chinese Academy of Sciences, Beijing 100190, P.R. China}
\abstract{Building upon known supersymmetric backgrounds, we derive novel half-BPS fermionic solutions in three-dimensional supergravity.
By virtue of an essential dependence on fermionic degrees of freedom, they possess no purely bosonic analogue.
In the Anti de Sitter case this notably includes nonsingular solutions for which the corresponding Chern-Simons gauge field $\mathcal{A}=\omega\pm e/L$ vanishes, providing access to configurations which are ordinarily singular in pure gravity.}

\notoc

\begin{document}
\maketitle
  
\vspace{8cm}
\section{Introduction}

By virtue of the symmetry they preserve, fully or fractionally supersymmetric solutions in supergravity possess special properties and utility, and accordingly, have over many years been thoroughly investigated and classified \cite{Tod:1983pm, Behrndt:1997ny, Gauntlett:2002nw, FigueroaO'Farrill:2002ft, Gauntlett:2002fz, Gutowski:2003rg, Bellorin:2005zc, Cacciatori:2007vn, Bellorin:2007yp, Gran:2008vx}.
However, despite this intense research effort it is nonetheless notable that the space of supersymmetric solutions is almost entirely unexplored in one direction; that is, when the fermions are non-trivial.

This of course needn't be the case.
Indeed the status of the gravitino as a gauge field suggests that it could play just as much of a non-trivial role as the graviphoton does in various supersymmetric solutions, such as the extremal Reissner-N{\"o}rdstrom black hole spacetimes \cite{Gibbons:1982fy}.
On a more prosaic level, fermion-sourced terms are in any case always present in supersymmetric theories, and to neglect them in all circumstances is perhaps something of a waste.

Whilst half-BPS instanton configurations for the gravitino field were presented in \cite{Houston:2016nbk}, more general constructions of supersymmetric fermionic solutions are nonetheless lacking.
This is the issue to which we presently direct our attention.

Sensible starting points are of course the familiar supersymmetric bosonic backgrounds, which we will augment with fermionic configurations preserving some or all of the underlying symmetry.
A priori this would seem to be possible only in certain restricted scenarios.
Given the existence of a Killing spinor $\chi$ we have
\begin{align}	
	\delta\psi
	=D\chi
	=0\stackrel{\psi\neq0}{\longrightarrow}
	\frac{1}{4}K^{ab}\gamma_{ab}\chi+\dots\,,
	\label{Killing spinor equation}
\end{align}
where $K^{ab}$ is the contortion one-form corresponding to the presence of torsion via 
\begin{align}
	T^a\equiv De^a
	=K^{ab}\wedge e_b
	\sim\overline\psi\wedge\gamma^a\psi\,,\quad
	de^a+\omega_\circ^{ab}\wedge e_b=0\,,
	\label{first Cartan structure equation}
\end{align}
and we have decomposed the full connection into curvature and contortion one-forms via $\omega^{ab}=\omega_\circ^{ab}+K^{ab}$\footnote{We work in `mostly plus' signature with $\left\{\gamma^a,\gamma^b\right\}=2\eta^{ab}$, using roman characters $\{a,b,\dots\}$ for frame indices, and greek characters $\{\alpha,\beta,\dots\}$ for spacetime indices. 
Antisymmetrisation is with unit weight, so that $\gamma^{a_1a_2\dots a_n}\equiv\gamma^{[a_1}\gamma^{a_2}\dots\gamma^{a_n]}=\frac{1}{n!}\left(\gamma^{a_1}\gamma^{a_2}\dots\gamma^{a_n}+\mathrm{permutations}\right)$ etc.
Our conventions follow \cite{Freedman:2012zz}.}.
The ellipsis indicates other possible fermionic terms arising from the possible gaugings and matter couplings of the specific supergravity theory in question.

Most obviously we could leverage the usual projection conditions associated to BPS states to engineer $\gamma_{ab}K^{ab}\chi=\gamma_{ab}K^{ab}P_\pm\chi=0$, such that \eqref{Killing spinor equation} still vanishes.
Alternatively, in the presence of matter couplings or gaugings we could instead arrange cancellation of the various fermionic terms amongst each other without requiring any projection conditions.

However, whilst these two possibilities certainly deserve further examination, they essentially correspond to privileged classes of fermionic excitations about familiar bosonic backgrounds.
An arguably more interestingly possibility would involve fermionic contributions cancelling against unwanted `bosonic' terms in the Killing spinor equations.
By thereby relying in an essential way upon fermionic fields, this would allow access to supersymmetric solutions which lack any purely bosonic counterpart.

\section{Supersymmetric, fermionic solutions}
To illustrate what is meant by this we may consider the simplest possible example of three-dimensional supergravity, defined by the Lagrangian
\begin{align}
	\mathcal{L}=2e_a\wedge R^{a}-\overline\psi\wedge D\psi\,,\quad
	R^a\equiv d\omega^{a}+\frac{1}{2}\epsilon^{abc}\omega_b\wedge \omega_c
	=\frac{1}{2}\epsilon^{abc}R_{bc}\,,
\end{align}
where for convenience we have a total connection
\begin{align}
	\omega^a=\omega_\circ^a+K^a+\frac{1}{L}e^a\,,\quad 
	K^a=\frac{1}{2}\epsilon^{abc}K_{bc}\,\,\, \mathrm{etc}\,,
	\label{total connection}
\end{align}
comprised respectively of curvature, contortion and cosmological constant terms.
The usual spinorial derivative can then be written
\begin{align}
	D\chi
	=\left(d+\frac{1}{4}\gamma_{ab}\left(\omega_\circ^{ab}+K^{ab}\right)-\frac{1}{2L}\gamma_ae^a\right)\chi
	=\left(d-\frac{1}{2}\gamma_a\omega^a\right)\chi\,,
	\label{Killing spinor equations}
\end{align}
where we have used the gamma matrix duality relations
\begin{align}
	\gamma^{abc}=-\epsilon^{abc}\,,\quad
	\gamma^{ab}=-\epsilon^{abc}\gamma_c\,,\quad
	\gamma^{a}=\frac{1}{2}\epsilon^{abc}\gamma_{bc}\,.
	\label{gamma matrix identities}
\end{align}
Assuming that there exists a complete set of Killing spinors satisfying $D\chi=0$, then
\begin{align}
	D^2\chi
	=-\frac{1}{2}\gamma_aR^a\chi
	=0\,,
	\label{Killing spinor integrability}
\end{align}
and the Einstein equations (incorporating torsion) are evidently satisfied.

The familiar bosonic backgrounds satisfying this requirement are of course flat and Anti de Sitter (AdS) space.
As explained in the introduction, we choose these backgrounds as our basis for supersymmetric fermionic solutions.
This being the case it apparently seems that $K^a\neq 0$ cannot be compatible with supersymmetry, as $D\chi$ will then no longer vanish as before.

However, given the presence of novel fermionic terms we are no longer obligated to use these backgrounds in their conventionally appropriate setting.
For example, using a flat metric in the context of `AdS supergravity' yields
\begin{align}
	\delta\psi
	=D\chi
	=-\frac{1}{2}\gamma_a\left(K^a+\frac{1}{L}e^a\right)\chi\,,
	\label{flat space gravitino variation}
\end{align}
where we make the reasonable assumption of constant Killing spinors.
If fermionic contributions can cancel against the `cosmological constant' term, then the Killing spinor and Einstein equations will again be satisfied as required. 
The usual scale $L$ may then be interpreted as the characteristic length associated to the torsion-induced twisting of reference frames under parallel transport, rather than the scale of background curvature.

\subsection{Minkowski space}

To engineer this in practice we firstly choose the gravitino
\begin{align}
	\psi=\gamma_ae^a\epsilon\,,\quad
	\overline\psi\equiv\psi^TC=-\overline\epsilon\gamma_ae^a\,,
	\quad d\epsilon=0\,,
\end{align}
following the conjugation conventions of \cite{Freedman:2012zz}.
Given the algebraic torsion constraint \footnote{\,As the algebraic torsion constraint is derived by varying the action with respect to $K^a$, unlike the Einstein and Rarita-Schwinger equations (which arise from varying $e^a$ and $\psi$ respectively) it is unmodified by the presence of a cosmological constant, there being no coupling between $K^a$ and such a term.}
\begin{align}
	T^a
	=de^a+\epsilon^{abc}\left(\omega_\circ{}_b+K_b\right)\wedge e_c
	=\frac{1}{4}\overline\psi\gamma^a\wedge\psi\,,
	\label{algebraic torsion constraint}
\end{align}
we may leverage \eqref{gamma matrix identities} and $\gamma^a\gamma^b\gamma^c=\gamma^{abc}+\gamma^a\eta^{bc}-\gamma^b\eta^{ca}+\gamma^c\eta^{ab}$ to give
\begin{align}
	\epsilon^{abc}K_{b}\wedge e_c
	=-\frac{1}{4}\left(\overline\epsilon\gamma^b\gamma^{a}\gamma^c\epsilon\right)e_b\wedge e_c
	=-\frac{1}{4}\epsilon^{abc}\left(\overline\epsilon\epsilon\right)e_b\wedge e_c\,.
\end{align}
Choosing the normalisation $\left(\overline\epsilon\epsilon\right)=4/L$, we have $K^{b}=-e^b/L$ and the `cosmological constant' term in \eqref{flat space gravitino variation} is cancelled as desired, ensuring that the Einstein equations are satisfied.
Furthermore, since $\omega^a$ then vanishes the Rarita-Schwinger equation is simply
\begin{align}
	D\psi
	=d\psi=0\,.
\end{align}

As our gravitino is non-trivial we must also consider the full supersymmetry variations
\begin{align}
	\delta\psi=D\chi=0\,,\quad
	\delta e^a
	\sim\overline\chi\gamma^a\psi
	=\left(\overline\chi\gamma^{ab}\epsilon\right)e_b+\left(\overline\chi\epsilon\right) e^a\,,
\end{align}
which appear to be non-vanishing.
However since $\chi$ and $\epsilon$ are both Killing, these bilinears correspond respectively to local Lorentz and scale transformations of the frame field. 
The former leave physical quantities invariant as required
\footnote{Indeed, under these transformations $\delta g_{\mu\nu}=e^a{}_{(\mu}\eta_{|ab|}\delta e^b{}_{\nu)}\sim \overline\chi\gamma_{(\mu\nu)}\epsilon=0\,.$},
whilst we may remove the latter via a projection condition, such as
\begin{align}
	P_\pm\equiv\frac{1}{2}\left(1\pm i\gamma_0\right)\,,\quad
	P_\pm\epsilon=\pm\epsilon\,,\quad
	P_\pm\chi=\mp\chi\,.
\end{align}
The bilinear $\left(\overline\chi\epsilon\right)$ then vanishes, and our solution is evidently half-BPS.
As torsion induces the twisting of reference frames under parallel transport, this is evidently sensible as we must choose between the two possible `helicities' at our disposal.

\subsection{Anti de Sitter space}

Naturally we can also apply this construction in the opposite, and likely more interesting, sense, where AdS space with fermionic support provides a basis for solutions in supergravity theories lacking a cosmological constant term in the action.
Broadly speaking, this amounts to the construction of a flat connection, or equivalently a parallelisation, where we eliminate curvature at the expense of introducing torsion.
Whilst this is not always possible for arbitrary manifolds, in the present context there are in principle no obstructions \cite{Figueroa-OFarrill:2003fkz}.

We work in the framework of $\mathcal{N}=\left(1,1\right)$ AdS supergravity, in the limit of vanishing cosmological constant. 
To be precise we mean this in the sense of working with the usual action with cosmological terms \cite{Achucarro:1987vz}, and only taking the vanishing limit at the level of the equations of motion and supersymmetry variations.
This ensures that the underlying $\left(1,1\right)$ structure induced by the $SO(2,2)=Sp\left(2,\mathbb{R}\right)\otimes Sp\left(2,\mathbb{R}\right)$ isometry of AdS${}_3$ does not degenerate into 
an inequivalent $\mathcal{N}=2$ Poincar\'e supergravity \cite{Lodato:2016alv}, as the physical $L$ contained in the metric nonetheless remains arbitrary.

Since our connection is flat the Rarita-Schwinger equation is simply $d\psi^I=0$, and locally we may always write $\psi^I=d\epsilon^I$ \footnote{We add an extra index $I$ to spinor fields to indicate which supergroup factor they correspond to.}.
In fact it will be opportune to identify $\epsilon^I$ with conventional AdS Killing spinors from the torsion-free context, the global existence of which can be assured.
Furthermore, because such Killing spinors generate global isometries they are inadmissible as (small) gauge parameters, ensuring the non-triviality of our result.

Following \cite{Freedman:2012zz} we use the AdS${}_3$ line element
\begin{align}
	ds^2=\exp\left(2r/L\right)\eta_{\mu\nu}dx^\mu dx^\nu+dr^2\,,\quad
	\mu,\nu\neq r\,,
\end{align}
such that $\epsilon^I$ must then satisfy
\begin{align}
	\left(\partial_r-\frac{1}{2L}\gamma_ae^a{}_r\right)\epsilon^I=0\,,\quad
	\left(\partial_\mu-\frac{1}{2L}\gamma_ae^a{}_\mu\left(1-\gamma_r\right)\right)\epsilon^I=0\,.
	\label{AdS Killing spinor conditions}
\end{align}
Noting the projection condition $P_-\equiv\left(1-\gamma_r\right)/2$, and the fact that $\overline{P_-\epsilon}=\overline\epsilon P_+$, we have
\begin{align}
	T^a
	=\frac{1}{4}\overline\psi^I\gamma^a\wedge\psi^I
	=\frac{1}{4L^2}\left(\overline\epsilon_+^I\epsilon_-^I\right)e^r\wedge e^a
	=\frac{1}{2L}\left(\overline\epsilon_+^I\epsilon_-^I\right)de^a\,,
	\label{AdS torsion}
\end{align}
where we have used $\left(\overline\epsilon_+^I\epsilon_-^I\right)=\left(\overline\epsilon_-^I\epsilon_+^I\right)$ and the gamma-matrix identities of the previous section.
Application of \eqref{AdS Killing spinor conditions} ensures that $d\left(\overline\epsilon_+^I\epsilon_-^I\right)=0$, and we can normalise this bilinear to $2L$, such that \eqref{algebraic torsion constraint} then implies $K^a=-\omega_\circ^a$.
In the absence of a cosmological constant term in the action the total connection $\omega^a$ then vanishes as required, and the Einstein equations are automatically satisfied.

As the connection is flat we again have a full complement of `Killing' spinors 
\footnote{Inverted commas distinguish between these and the Killing spinors of the purely bosonic background.}, whilst we must contend with the frame field transformation
\begin{align}
	\delta e^a
	\sim\overline\chi^I\gamma^a\psi^I
	=\overline\chi^I\left(\gamma^{ar}+\delta^{ar}\right)\epsilon^Ie^r+\overline\chi^I\left(\gamma^{ab}+\eta^{ab}\right)\epsilon^I_-e^b{}_\mu\,.
\end{align}
Whilst the antisymmetric terms have no effect on the metric variation, as in the flat case discussed previously, the symmetric terms are problematic as if $\epsilon^I=\epsilon^I_-$ then by \eqref{AdS torsion} $T^a$ must vanish. 
This precludes the use of a projection condition to remove them.
We are however free to choose $\psi^I=\left\{d\epsilon,0\right\}$ or vice-versa, such that $\left(0,1\right)$ or $\left(1,0\right)$ invariance is manifest.
Alternatively we may identify $\psi^1=-\psi^2$, such that invariance occurs for $\chi^1=\chi^2$, thereby reducing the number of independent parameters contained in the Killing spinors by half.
This ensures that $\delta e^a$ vanishes, and our solution is half-BPS
\footnote{Strictly speaking we could use just one of the four available Killing spinors, and find a 3/4 BPS result. 
We will elaborate on this more fully in a forthcoming publication.}.

\section{Discussion \& conclusions}

We have constructed novel half-BPS fermionic solutions in supergravity, which, by virtue of the essential role played by fermionic fields, possess no purely bosonic equivalent.
Although given the simplicity of three-dimensional gravity these are uncomplicated examples in many respects, their existence nonetheless suggests a way forward. 
Familiar supersymmetric solutions may be imported into an `inappropriate' supergravity theory, with the now-unwanted `bosonic' terms in the Killing spinor equations then cancelled via fermionic contributions.

Whilst this approach may independently be of interest, the solutions found may also carry significance for the study of gravity in three dimensions.
As is well established, $(1,1)$ (super)gravity in three dimensions admits an equivalent description in terms of an $OSp\left(1|2;\mathbb{R}\right)\otimes OSp\left(1|2;\mathbb{R}\right)$ Chern-Simons gauge theory with two gauge fields
\begin{align}
	\mathcal{A}^I=A^a_I J_a^I+\psi^I Q_I\,,\quad
	A^a_I=\omega^a+\left(-1\right)^I e^a/\tilde L\,,\quad
	I=1,2\,,
\end{align}
where $J_I^a$ and $Q_I$ span a copy of the $OSp\left(1|2;\mathbb{R}\right)$ algebra for each value of $I$ respectively and $\tilde L$ corresponds to the cosmological constant term in the action, rather than necessarily the metric.
As emphasised in \cite{Witten:2007kt}, this equivalence is somewhat incomplete in the only background where a quantum theory of gravity is likely to make sense; AdS space.
This is in part due to the absence of the trivial $\mathcal{A}^I=0$ solutions on the gravity side, owing to the loss of metric invertibility for $e^a=0$.

However as we have no need for cosmological constant terms, and as a result find solutions where both $\omega^a$ and one of our gravitini vanish, AdS solutions for which e.g. $\mathcal{A}^1=0$ then do exist in supergravity by virtue of the parallelising effect of non-trivial fermions.
Various other discrepancies between the two descriptions still exist \cite{Witten:2007kt}, but it would nonetheless be interesting to explore the role fermionic configurations such as these have in the quantisation of (super)gravity in three dimensions, and indeed in the context of the AdS${}_3$/CFT${}_2$ correspondence 
\footnote{Generalisation to e.g. extremal BTZ spacetimes, noting the purely gravitational result of \cite{Alvarez:2014uda}, is presently underway.}.

It is furthermore interesting to observe that, as discussed in \cite{Witten:1988hc}, since the Riemann curvature relies on the inverse metric, configurations for which $e^a$ or equivalently $A^I$ vanish somewhere must be regarded as singular in classical general relativity. 
By including fermionic contributions in the connection we have however incorporated solutions of this type in a non-singular fashion. 
It remains an open question as to whether this phenomenon may be exploited more generally.
Generalisation of these solutions to higher dimensions and more sophisticated supergravity theories also remains.

\acknowledgments

We enjoy fellowship support from the Chinese Academy of Sciences and the Royal Society.

\end{document}